# THERMODYNAMICS
# AND THE ORIGIN OF LIFE

**Gerald E. Marsh**
**Argonne National Laboratory (Retired)**
**gemarsh@uchicago.edu**

## ABSTRACT

Modern developments in nonequilibrium thermodynamics have significant implications for the origins of life. The reasons for this are closely related to a generalized version of the second law of thermodynamics recently found for entropy production during irreversible evolution of a given system such as self-replicating RNA. This paper is intended to serve as an introduction to these developments.

# Introduction

Statistical physics and non-equilibrium thermodynamics, and their relation to biological systems, have long been considered a part of the physics corpus albeit not exclusively. See for example the book *Statistical Physics* by Landau and Lifshitz [1]. And from a more basic and philosophic point of view, one can date the major contributions of physicists to biology back to at least the last fifty years of the 20th century, and in particular to Erwin Schrödinger's book *What is Life*? [2]. Many research papers relating physics to the origins of life and to biology in general were also published during this period.

This paper is intended to be a limited introduction to the importance of non-equilibrium thermodynamics and a generalized form of the second law, which have great significance for the origin of life. It is not intended to be a comprehensive review of the literature, and the reader is encouraged to consult the original publications for additional detail and context. Nonetheless, a brief but limited historical review may be helpful. It is hoped that this, and what follows, will facilitate entry into this literature.

## Historical Review

In the mid-1950s, nonequilibrium thermodynamics was developed for application to systems dependent on the flow of entropy from the external environment. It was found that nonlinear phenomena were more the rule than the exception in natural systems. The paper cited in the next paragraph by Prigogine and Nicolis provides an excellent introduction.

Consider the Boltzmann factor $\exp(-E/kT)$. Energy becomes the dominant factor at low temperature, where the entropy contribution to the free energy is small, so that the system adopts a configuration that minimizes the potential energy. Prigogine and Nicolis [3] termed this the "Boltzmann order principle". If the system is driven far from equilibrium, a new ordering principle, which they called "order through fluctuations", can appear that gives rise to "ordered, highly-co-operative structures". They then noted that this phenomenon also appears in far from equilibrium chemical processes relevant to biology.

Long ago it was shown by Einstein that the probability of fluctuations around equilibrium in an isolated system was given by $P \propto \exp(\Delta S/k)$, where $\Delta S$ is the entropy change, which is less than zero for fluctuations. Later it was shown that Einstein's fluctuation theory could be extended to non-equilibrium states.

In the early 1970s Glansdorff and Prigogine [4], as well as others [5,6,7], extended the concept of macroscopic nonequilibrium thermodynamics into the nonlinear regime far from equilibrium. In doing so they assumed that a local entropy can be defined that depends on the same independent variables that are used in the equilibrium case. They were able to show that stable states exist for dissipative systems far from equilibrium. These were termed by Prigogine "dissipative structures"[†], which appeared far from equilibrium when large external forces led the system into the nonlinear regime; these structures increase the dissipation of the applied force.

In 2004, Andrieux and Gaspard [8] were able to derive the Onsager [9] and higher-order reciprocity relations[††] from a fluctuation theorem for nonequilibrium reactions. The time evolution of such systems is assumed to be a Markovian stochastic process, where the probability distribution of future states only depends on the present state. Living systems, however, cannot be represented by Markovian statistical approaches. Instead, the Glansdorff-Prigogine approach, based on the change in entropy production due to changes in the generalized forces, is thought to be more appropriate. Life evolved over stationary states due to the inherent non-linearity between generalized flows and forces. Because of the dependence on initial conditions, and subsequent perturbations, the process is distinctly non-Markovian.

Michaelian [10] has given a comparison of the usual statistical mechanical ideas with nonequilibrium thermodynamics and dissipative structuring; and has also introduced the idea that the molecules necessary for life could be considered to be dissipative structures driven to self-organize by UV light, thereby using the formalism of nonequilibrium thermodynamics to address the origin of life and its evolution [11,12,13].

---

[†] In their book, *Self-Organization in Non-Equilibrium Systems* (John Wiley & Sons, Inc. 1977), Nicolis and Prigogine state ". . . we have found that the distance from equilibrium and the nonlinearity may both be sources of order capable of driving the system to an ordered configuration. . . . we call the ordered configurations that emerge beyond instability of the thermodynamic branch the *dissipative structures*."

[††] In a non-equilibrium system, there could be flows of energy and mass where the flows are coupled. This can be represented as $J_\alpha = \sum_\beta L_{\alpha\beta} \nabla F_\beta$, where the $J_\alpha$ represent the flows and $F_\beta$ the "forces" producing the flows. Onsager showed that $L_{\alpha\beta}$ (the matrix of transport coefficients) is positive semi-definite and symmetric—except in cases where time-reversal symmetry is broken.

**Origin of Life**

It is generally believed that life began with the evolution of self-replicating polynucleotides. Ever since the paper by Gilbert [14] that discussed the possibility that catalytic RNA enzymes or ribozymes could be involved in the evolution of life, and incidentally coined the term the "RNA world", RNA has been the favorite molecule. Gilbert's *News and Views* paper addressed the previous week's *News and Views* paper by Westheimer [15]. The original discovery of the enzymatic activities of RNA was by Cech [16]. In 2009, Lincoln and Joyce [17] showed the self-sustained replication of two ribozymes that catalyze each other's synthesis. These cross-replicating ribozymes grew exponentially in the absence of proteins or other biological materials.

For an RNA world to exist, there must be some way for RNA to form on the early earth. How these nucleotides could have formed concurrently under the geophysical constraints of the early Earth was an unsolved chemical mystery before Becker, et al. [18] published their paper "Unified prebiotic plausible synthesis of pyrimidine and purine RNA ribonucleotides" in 2019. RNA molecules are composed of purine and pyrimidine nucleotides. Becker, et al. found a reaction network under which both nucleotides could simultaneously form when driven by wet-dry cycles. The chemical reactions involved are very complex, but Hud and Fialho [19] gave a relatively simple summary of them, which can be outlined as shown in Figure 1.

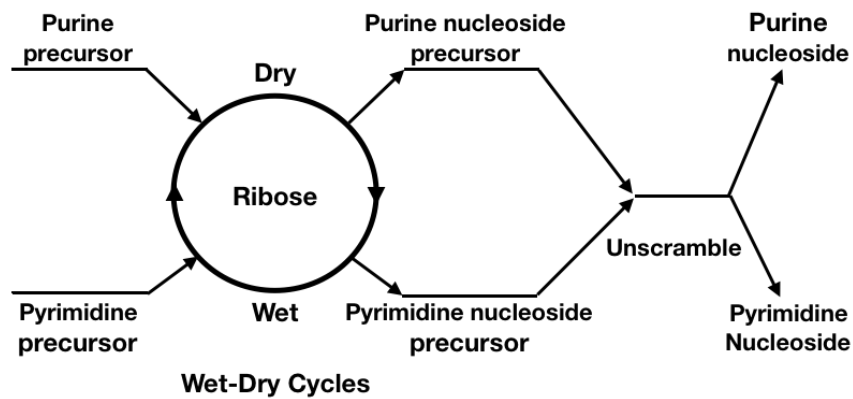

Figure 1. A simplified reaction network for the simultaneous formation of purine and pyrimidine nucleosides. When a mixture of the sugar ribose and non-natural nucleobases are subjected to wet-dry cycles, intermediate molecules are formed that can be converted to the natural nucleosides, which when phosphorylated give the natural nucleotides. For a discussion of what "scrambled" means see the article by Hud and Fialho, which supplied the basis for this figure. They used the term to designate the precursor formed by Becker, et al., which they called a "scrambled" pyrimidine nucleobase.

The geochemical conditions in the current work of Becker, et al. is compatible with their previous synthesis of the purine nucleosides [20]. The wet-dry cycles allow the dried reactants to coalesce into a concentrated state where their joining through covalent bond formation with the release of water molecules becomes thermodynamically favorable. What "thermodynamically favorable" means is the subject of much of what follows.

From a simplistic point of view, life appears to violate the second law of thermodynamics since ordered life forms have a lower entropy than their precursors. Of course, this apparent violation vanishes when one realizes that life processes draw upon the free energy from the surroundings so that total entropy increases. It has been known for some time that matter can spontaneously organize itself if the Gibbs free energy for the process is negative or if an external source of "activation energy" is provided. If this absorbed energy is dissipated after overcoming the activation barrier it is no longer available to drive the reverse process. This irreversible phenomenon is called "driven self-assembly". Explaining how this works for a macroscopic system is facilitated by the introduction of a generalized form of the second law of thermodynamics.

A short Appendix discusses some of the terminology, definitions, and confusions associated with thermodynamic quantities.

## Nonequilibrium States

A system may be maintained in a nonequilibrium state by a flow of energy. If the state is time independent, macroscopic observables will have constant nonequilibrium values. The example often given is a constant electrical current through a resistor with a steady rate of heat generation. Such dissipative structures can be formed and maintained by irreversible processes that continuously increase entropy [21,22]. In a linear regime, small deviations in the forces driving a flow (such as the heat flow on the resistor example) will lead to the flow being a linear function of the forces driving the flow. But a system that is not in thermodynamic equilibrium need not be in a stationary, time independent state since systems that are far from equilibrium can become dependent on nonlinear empirical relationships; i.e., phenomenological models.

Systems of identical non-interacting components in thermodynamic equilibrium can be described by the Boltzmann distribution (also known as the Gibbs distribution), which gives the equilibrium probability distribution of different energy states of a system as a function of the state's energy and the temperature of the system. It has the general form $P_i \propto \exp\left[-E_i/k_B T\right]$, where $k_B$ is the Boltzmann constant, $P_i$ is the probability that the system is in the state $i$ and $E_i$ is the energy of the state. Thus, if the states $j$ and $k$ have energies $E_i$ and $E_j$, the relative probability is

$$\frac{P_i}{P_j} = \exp\left[\frac{E_j - E_i}{k_B T}\right].$$

(1)

Moreover, systems in thermodynamic equilibrium satisfy the principle of detailed balance, which is equivalent to microscopic reversibility.

If the system is driven far from equilibrium into the nonlinear regime, the Boltzmann distribution is no longer valid since the thermodynamic flows are no longer linear functions of the thermodynamic forces. Far from equilibrium states can evolve into one of many new, highly organized states also known as dissipative structures.

## Microstates and Reversibility

The statistical behavior of nonequilibrium systems require the introduction of comparisons between the dynamical trajectories of the components of the system rather than the local properties of individual microstates at one moment of time [23]. If the dynamics of a system are stochastic and Markovian (meaning a sequence of events where the probability of each event depends only on the state of the previous event), one can require that the dynamics follow the microscopically reversible condition [24]

$$\frac{P[x(+t)|\lambda(+t)]}{P[\overline{x}(-t)|\overline{\lambda}(-t)]} = exp\{-\beta \, Q[x(+t), \lambda(+t)]\},$$

(2)

where $\beta = 1/k_B T$, the state of the system is given by the function $x$, representing all dynamical uncontrolled degrees of freedom, while $\lambda$ is a controlled time-dependent parameter, which depends on the type of system being considered.

In Eq. (2), $P[x(+t)\,|\,\lambda(+t)]$ is the probability of following the path $x(+t)$ through phase space and the denominator is the corresponding time-reversed path. This notation for the time-reversed path is a consequence of changing the time origin so that $t \in \{-\tau, \tau\}$, where $\tau$ could be infinite. The overbar indicates that quantities odd under time reversal also change sign. $Q$ is the amount of energy in the form of heat transferred to the system from the heat bath. $Q$ is a function of the phase space path and odd under time reversal; i.e., $Q[x(+t), \lambda(+t)] = -Q[\overline{x}(-t), \overline{\lambda}(-t)]$.

Eq. (2) is microscopically reversible; it relates the probability of a given path to its reverse path. Note that this is not the same as the principle of detailed balance, which refers to the probabilities of changing states independent of path. This distinction is important because Eq. (2) holds when the system is driven by an external time varying force field. England [25] has used this equation to derive a generalization of the second law of thermodynamics that is important for many far from equilibrium thermodynamic systems. It applies to the macroscopic transition between complex course-grained states.

## Dissipative Adaptation

To be consistent with England's notation and make it easier to read his paper, we now change notation so that Eq. (2) can be rewritten as

$$\frac{\pi[\gamma]}{\pi^*(\gamma^*)} = \; exp\left[\frac{\Delta Q(\gamma)}{kT}\right],$$

(3)

where $\gamma$ is a microtrajectory and the "$*$" indicates time-reversed. Here, it is assumed that $\Delta Q(\gamma)$ is the sum of the internal energy change of the system when traversing the path $\gamma$ and the work applied to the system by an external field; i.e., $\Delta Q(\gamma) = \Delta E + W$.

Let $i$, $j$, and $k$ represent different possible configurations of a system composed of distinct components (such as particles). Equation (3) then becomes

$$\frac{\pi[i{\to}j]}{\pi^*(j^*{\to}i^*)} = \; exp\left[\frac{\Delta Q(i{\to}j)}{kT}\right].$$

(4)

Using Eq. (4) also for the trajectory $i{\to}k$ and taking the ratio of the equations for the transition $i{\to}j$ and $i{\to}k$, that is,

$$\frac{\dfrac{\pi[i{\to}j]}{\pi^*(j^*{\to}i^*)}}{\dfrac{\pi[i{\to}k]}{\pi^*(k^*{\to}i^*)}} = \frac{exp\left[\dfrac{\Delta Q(i{\to}j)}{kT}\right]}{exp\left[\dfrac{\Delta Q(i{\to}k)}{kT}\right]},$$

(5)

with the substitution $\Delta Q(\gamma) = \Delta E + W$, after some algebra Eq. (5) will yield

$$\frac{\pi[i{\to}j]}{\pi[i{\to}k]} = exp\left[\frac{-\Delta E(j{\to}k)}{kT}\right]\frac{\pi^*(j^*{\to}i^*)}{\pi^*(k^*{\to}i^*)}\frac{exp\left[-\dfrac{W(i{\to}k)}{kT}\right]}{exp\left[-\dfrac{W(i{\to}j)}{kT}\right]}.$$

(6)

The first term on the right-hand-side of Eq. (6) comes from the term $exp\left[\frac{\Delta E(i \rightarrow j) - \Delta E(i \rightarrow k)}{kT}\right]$ encountered when doing the algebra. The energy level relationships of Eq. (6) are shown in the figure below

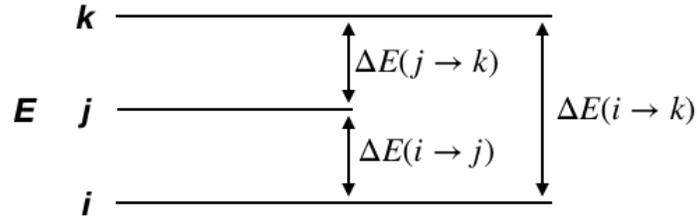

From this figure one can see that $\Delta E(i{\rightarrow}j) - \Delta E(i{\rightarrow}k) = \Delta E(j{\rightarrow}k)$.

To explain the concept of dissipative adaptation, England assumes that the states have the same energy so that the first term on the right-hand side of Eq. (6) is unity, and averages over all microtrajectories with fixed endpoints. With these assumptions, Eq. (6) becomes

$$\frac{\pi[i{\rightarrow}j]}{\pi[i{\rightarrow}k]} = \frac{\pi^*(j^*{\rightarrow}i^*)}{\pi^*(k^*{\rightarrow}i^*)} \frac{\left\langle exp -\dfrac{W(i{\rightarrow}k)}{kT}\right\rangle}{\left\langle exp -\dfrac{W(i{\rightarrow}j)}{kT}\right\rangle}.$$

(7)

The brackets $\langle \ldots \rangle$ indicate the average over microtrajectories. Note that even though the states now have the same energy, the left-hand side of Eq. (7) could differ from unity since not all states of the system are equally accessible in a finite time.

For different oscillatory external forces, known as the "drive", different configurations of the system will absorb work from the external forces at different rates. A given system configuration could then surmount "activation barriers" to transition to states that would not be accessible through thermal fluctuations alone. This is shown in Fig. 2 for the general case where the energies of the states are not identical.

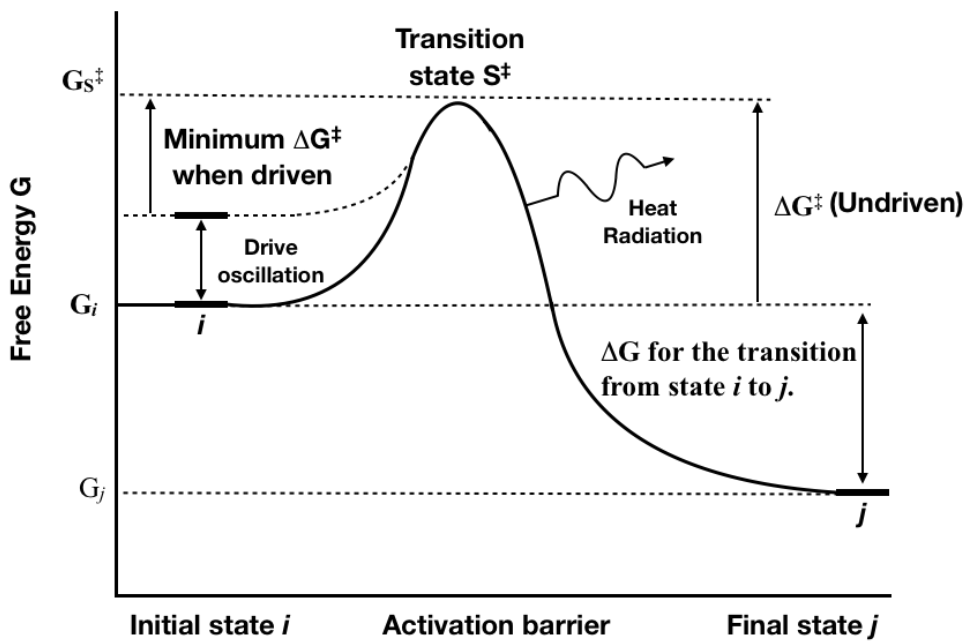

Figure 2.  Transition through a system configuration activation barrier.  $\Delta G^{\ddagger} = G_S^{\ddagger} - G_i$, where $G_i$ oscillates when driven.  For the case where the energies of the states $i$ and $j$ are equal, $G_i = G_j$.  The energy absorbed from the drive is radiated as heat in the transition from $S^{\ddagger}$ to state $j$.  The symbol used to indicate heat radiation corresponds to the full spectrum of the heat being radiated.  This figure should be compared to Fig. A1 in the Appendix for the activation energy of a chemical reaction.

The point of all this is that some of the randomly changing system configurations will be better able to absorb work from the drive than others and this leads to a mostly one directional change in configurations—due to the loss of the radiated heat energy—to those better able to absorb and dissipate the energy absorbed from the drive.  As put by England, the structure will appear to self-organize into a state that is well adapted to the environmental conditions set by the drive.  This, he calls the phenomenon of "dissipative adaptation".

On the other hand, the concept of "dissipative adaptation" is still somewhat controversial with regard to finding a clear link between Eq. (7) and actual physical systems.  Equation (7) tells us that trajectories that absorb work will be favored over their reversed trajectories but this does not necessarily imply that such a trajectory "will be kinetically favored over completely different routes through configuration space that start from the same point."[†].  Kachman, Owen, and England have shown, using a simulated "toy chemical" reaction, that "elevated work absorption during the system's history has enabled irreversible configurational change, except in this case, the outcome states are less capable of absorbing work than their predecessors." [26]

---

† A point made by an anonymous referee.

In this vein, Kondepudi and Prigogine in their 2015 book *Modern Thermodynamics* (Ref. 21) introduce the concept of "structural instability" that can occur when a new chemical species is introduced into a nonequilibrium chemical system that could destabilize the system so that it evolves into a new state. They liken this to Darwinian evolution at the molecular level. Thus, in their words, "we see instability, fluctuation and evolution to organized states as a general nonequilibrium process whose most spectacular manifestation is the evolution of life."

### Macroscopic Irreversibility

The relationship between microscopic irreversibility and entropy production has thermodynamic effects on far from equilibrium macroscopic processes and in particular for biological self-replication (to be discussed in the next section). England derived a generalized form of the second law of thermodynamics that can be written as

$$\beta \langle \Delta Q \rangle_{I \to II} + ln\left[\frac{\pi(II \to I)}{\pi(I \to II)}\right] + \Delta S_{int} \geq 0.$$

(8)

The derivation of this equation begins with Eq. (3) with $\gamma \to x(t)$, where $0 \leq t \leq \tau$. Taking the natural logarithm of the resulting equation gives

$$\beta \Delta Q = ln\left[\frac{\pi[x(t)]}{\pi[x(\tau - t)]}\right],$$

(9)

an equation concerned with microscopic irreversibility.

Suppose there is a coarse-grained observable, *I*, which can be associated with a probability distribution $p(i|I)$, the probability that it is in a microstate $i$. If there is a second course-grained observation of the system after a time interval $\tau$ designated by *II*, $p(j|II)$ is defined as the probability that the macrostate *II* (that evolved from macrostate *I* after a period of time $\tau$) is in the microstate *j*. Note that the macrostates I and II are ensembles of paths.

Crucially, these probability functions allow a macroscopic definition of irreversibility:

$$\pi(I\rightarrow II) = \int_{II} dj \int_{I} di\, p(i\,|\,I)\pi(i\rightarrow j)$$

$$\pi(II\rightarrow I) = \int_{I} di \int_{II} dj\, p(j\,|\,II)\pi(j\rightarrow i).$$

(10)

Similar to what was done in Eq. (5), taking the ratio of $\frac{\pi(II\rightarrow I)}{\pi(I\rightarrow II)}$ gives, after some algebra,

$$\frac{\pi(II\rightarrow I)}{\pi(I\rightarrow II)} = \left\langle \left\langle e^{-\beta\,\Delta Q_{ij}}\right\rangle_{i\rightarrow j} \middle/ \exp\left[ln\left[\frac{p(i|I)}{p(j|II)}\right]\right]\right\rangle_{I\rightarrow II}.$$

(11)

The first averaging bracket on the right-hand side is the average of all paths from $i \in I$ to $j \in II$, each path being weighted by its likelihood (the second averaging bracket).

The next step is to introduce the Shannon entropy $S = -\sum_i p_i\, ln p_i$ so that an expression for the internal entropy change for the transition between the ensembles *I→II* can be written. England uses units such that the Boltzmann constant is unity. This results in $\Delta S_{int} = S_{II} - S_I$ and after some additional algebra he obtains the generalization of the second law of thermodynamics given by Eq. (8); that is,

$$\beta\langle\Delta Q\rangle_{I\rightarrow II} + ln\left[\frac{\pi(II\rightarrow I)}{\pi(I\rightarrow II)}\right] + \Delta S_{int} \geq 0.$$

(12)

The first term in Eq. (12) is the entropy change of the heat bath and the entropy generated by second term vanishes if $\pi(II\rightarrow I) = \pi(I\rightarrow II)$, which would result in the usual second law of thermodynamics where $\beta\langle\Delta Q\rangle_{I\rightarrow II} + \Delta S_{int} \geq 0$ because the average entropy change of the universe must be greater than or equal to zero.

One might question whether the use of the Shannon entropy is legitimate. The Boltzmann distribution implies the usual thermodynamic definition of entropy. The Gibbs-Shannon entropy given by $S = -k_B \sum_i p_i \, ln p_i$ is equivalent to the thermodynamic definition of entropy only for what is known as the generalized Boltzmann distribution [27], which is valid for all Markovian systems even those not in thermodynamic equilibrium. The generalized Boltzmann distribution itself was defined by Lin [28] using an analogy based on electronics to give an explanation of the concept. Given a thermodynamic system with $m + n$ generalized forces and coordinates, Xiang, et al. write the probability density function $Pr(\vec{\omega})$ of the microstate $\vec{\omega}$ for the generalized Boltzmann distribution as

$$Pr(\vec{\omega}) \propto exp\left[\sum_{\eta=1}^{n} \frac{X_\eta x_\eta^{(\vec{\omega})}}{k_B T} - \frac{E^{(\vec{\omega})}}{k_B T}\right],$$

(13)

where the $X_\eta$ are generalized forces and $E$ and $x_\eta$ are random variables, and Xiang, et al. use the vector notation to designate a microstate, as in $\vec{\omega}$. If the generalized forces vanish, Eq. (13) reduces to the usual Boltzmann distribution.

The work of Xiang, et al. tells us is that for stable, non-equilibrium steady state systems, the entropy cannot be described by the Gibbs-Shannon entropy unless their distribution of states is given by the generalized Boltzmann distribution given in Eq. (13).

Since the systems considered above are Markovian, the use of the Shannon entropy there is indeed legitimate.

## Replicating Systems and the Generalized Second Law of Thermodynamics

England introduced the idea of a generalized second law of thermodynamics. Some believe this is unnecessary and that for open systems one can use the traditional approach by including the internal production of entropy due to irreversible processes and external flows of entropy into or out of the system. Nonetheless, the concept of a generalized second law is interesting and for that reason, if no other, is discussed here.

England has used self-replicating systems to illustrate the use of his generalized second law. Consider $n(t = 0) \gg 1$ for $n$ self-replicating molecules at an inverse temperature $\beta$. They would have an exponential growth given by

$$n(t) = n(0)e^{(g - \delta)t},$$

(14)

where $g$ determines the growth rate and $\delta$ the decay rate.

The probability that in a time $dt$ one particular replicator associated with $\pi(I{\rightarrow}II)$ would replicate would be given by $gdt$ and its decay probability $\pi(II{\rightarrow}I)$ would be $\delta\, dt$. Imposing the generalized second law of thermodynamics yields,

$$\Delta S_{tot} = \beta \Delta q + \Delta S_{int} \geq ln\left[\frac{g}{\delta}\right].$$

(15)

Note that if $g > \delta$, so that there is net growth, the total entropy associated with self-replication will have a positive lower bound.

This can be seen by setting $n(0) = 1$ in Eq. (14) and finding that the doubling time is proportional to $1/(g - \delta)$. Since the only requirement is that $g > \delta$, the doubling time can be made arbitrarily short. On the other hand, a very small value of for $\Delta S_{tot}$ can be obtained by making the difference between $g$ and $\delta$ very small, but $\Delta S_{tot}$ will nonetheless be positive. It is equal to zero only for $g = \delta$; i.e., when there is no net growth.

Two conclusions are readily apparent: (1) The growth rate of a self-replicator depends on its internal entropy ($\Delta S_{int}$), its durability ($1/\delta$), and the heat ($\Delta q$) dissipated into the surrounding heat bath in the process of replication; and (2), heat must be generated from energy stored in the reactants or work done on the system by a time varying external driving field.

## Implications for the Origin of Life

1.

Using the RNA molecule again as an example, how purine and pyrimidine nucleosides could have formed together under early Earth geophysical constraints was, as mentioned in the Introduction, until recently an unsolved chemical problem. From a global perspective, the resolution of this problem involved far from equilibrium thermodynamics—in this case where the system was driven by wet-dry cycles; it is this external forcing that made the formation of these nucleosides thermodynamically favorable.

A very interesting mechanism has been proposed by K. Michaelian (Ref 12) for the replication of RNA and DNA molecules without the need for enzymes. He proposed that UV light dissipation and diurnal temperature cycling of the Archean sea-surface can lead to such replication. It is based on what Michaelian calls a non-equilibrium thermodynamic imperative for producing RNA and DNA due to the great entropy producing potential of these molecules under the initial conditions of the primitive Earth.

While the idea that far from equilibrium thermodynamics is fundamental to the origin of life remains somewhat controversial, the case is quite strong for this point of view. There is also a simulated toy chemical model whose behavior is consistent with the idea of far from equilibrium self-organization [29].

General chemistry tells us that the free energy change in a reaction is governed by the equilibrium constant. Furthermore, while the reaction rate can be changed by enzymes which do not alter the equilibrium, the change in free energy is independent of the path or the molecular mechanism of the transformation. It is clear that this is not the case for nonequilibrium thermodynamics "whose most spectacular manifestation is the evolution of life".

Our Earth formed some 4.5 billion years ago and there is strong evidence that life appeared 700 million to 1 billion years later; that is, about 3.8 billion years ago. Far from equilibrium nonlinear thermodynamics in the presence of external drives could help explain this very rapid origin of life. Because the length of the day was only around seven or so hours long 3.8 billion years ago, the tides would be far higher than today and there would be strong diurnal forces that could play the role of an external drive including ultraviolet radiation, which would be far more intense than today, as pointed out by Michaelian, since there was little if any oxygen in the atmosphere and therefore no ozone layer to block the UV.

Life appears as a process ultimately founded on the ability of matter, governed by the principles of quantum mechanics, to form the molecules need for life to exist. These molecules combine, overcoming the limits normally imposed by both unfavorable free energy constraints and activation energies, because of the properties of driven nonlinear thermodynamic systems. Most stars have planets, and those with earthlike planets are all very likely to have life due to the reduction in the activation energy for the formation of complex biomolecules arising from driven, far from equilibrium nonlinear thermodynamics.

## Acknowledgement

I would like to thank Theodore L. Steck, M.D. for his many helpful suggestions and corrections. I would also like to thank an anonymous referee for careful readings of the manuscript and the suggestion that it lacked an historical context as well as several appropriate references. The Historical Review section, following the Introduction, is based on the referee's précis of the history.

# APPENDIX

There is much confusion in the literature about the term "energy" in physics, biology, and chemistry. The definitions here are from the International Union of Pure and Applied Chemistry (IUPAC). They define:

$U$ = the internal energy, $U = Q + W$, where $Q$ designates heat and $W$ work. In parts of this paper, to be consistent with the referenced literature, the internal energy is designated as $E$ and its change $\Delta E$.

$A$ = Helmholtz energy function: $A = U - TS$, where $S$ is the entropy and $T$ the temperature. Note that in the earlier literature $A$ was called the "free energy".

$H$ = enthalpy: $H = U + PV$, where $P$ is the pressure and $V$ the volume. $\Delta H$ is the heat brought to a system at constant pressure.

$G$ = Gibbs energy function: $G = H - TS$. Formerly called the "free energy" or "free enthalpy". It is the reversible useful work performed at constant temperature and pressure.

An additional confusion is that in physics the "free energy" is the generally the Helmholtz energy and in biology and chemistry it is the Gibbs free energy. In particular, in biochemistry, the Gibbs free energy is defined as

$$\Delta G = \Delta G^0 + RT \, ln\frac{[C][D]}{[A][B]},$$

where, $A + B \leftrightharpoons C + D$, and [$A$, $B$, $C$ or $D$] are the molar concentrations (the "activities") of the reactants. $\Delta G^0$ is the Gibbs free energy for this reaction under "standard conditions", meaning when $A$, $B$, $C$, $D$ are present at a concentration of 1.0 M, and the square brackets indicate concentrations.

In biochemistry, it is generally assumed that the initial and final states of a reaction are equilibrium states so that $\Delta G = G_{final} - G_{initial}$. $\Delta G$ is independent of the path or molecular mechanism of the transformation. This is not the case for the nonequilibrium states discussed in the body of this paper. A negative $\Delta G$ implies that the reaction can occur spontaneously. The rate of a reaction depends on the "free energy of activation" $\Delta G^{\ddagger}$ as shown in the figure below.

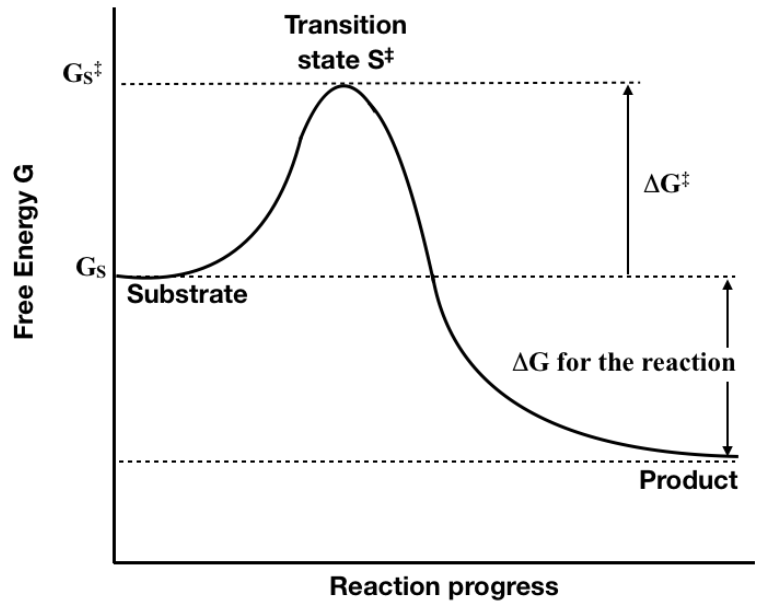

Figure A1. The transition state S$^{\ddagger}$ has a higher energy than either the substrate $S$ or product. The Gibbs free energy of activation, or "activation energy", is $G^{\ddagger} = G_{S^{\ddagger}} - G_S$, where $G_{S^{\ddagger}}$ is the Gibbs free energy of the transition state S$^{\ddagger}$, and $G_S$ is the free energy of the substrate.